# Charge-polarized superconducting state emerging in a superatomic antipolar metal

Shuya Xing[1,5,6,+,*], Zhongxu Wei[2,+], Xu Chen[2,*], Junming Zhang[2], Zhenyu Yuan[2], Jiali Zhao[2], Feng Jin[2], Tao Sun[2], Huifen Ren[2], Minjie Cui[2], Hong Chang[1,5,6], Tianping Ying[2], Jiangang Guo[2], Hechang Lei[3,4], Shifeng Zhao[1,5,6], Wenping Zhou[1,5,6], Xinqi Li[1,5], Tian Qian[2], Wei Ji[3,4,*], and Zhihai Cheng[3,4,*]

[1]*Research Center for Quantum Physics and Technologies, Inner Mongolia University, Hohhot 010021, China*

[2]*Beijing National Laboratory for Condensed Matter Physics, Institute of Physics, Chinese Academy of Sciences, Beijing 100190, China*

[3]*Beijing Key Laboratory of Optoelectronic Functional Materials & Micro-Nano Devices, School of Physics, Renmin University of China, Beijing 100872, China*

[4]*Key Laboratory of Quantum State Construction and Manipulation (Ministry of Education), Renmin University of China, Beijing, 100872, China*

[5]*School of Physical Science and Technology, Inner Mongolia University, Hohhot 010021, China*

[6]*Inner Mongolia Key Laboratory of Microscale Physics and Atom Innovation, Inner Mongolia University, Hohhot 010021, China*

**Abstract:** The simultaneous presence of polarity and metallicity or superconductivity in a material signifies the exotic polar metallic or superconducting (SC) state, while such materials are extremely rare due to their exclusive nature. Recently, the interweaved CDW and antipolar charge orders have been discovered in a metallic superatomic crystal of $Au_6Te_{12}Se_8$ (ATS), while their interplay and competition with the following emergent SC state remains elusive. Here, we report a further experimental investigation of the SC state emerged from the preformed CDW and antipolar order states using scanning tunneling microscopy/spectroscopy in combination with transport and Raman measurements. The temperature-dependent pre-formation and condensation of Cooper pairs are experimentally identified. The pre-existent CDW is gradually suppressed by the preformed Cooper pairs, and then the antipolar charge order is spatially suppressed into a ferrielectric-like polar order by the condensed Cooper pairs of SC state. The exotic charge-polarized superconducting state is discovered in the polar metal of ATS, suggesting a valuable platform for the exploration of intriguing polar superconducting properties.

[+] These authors contributed equally: Shuya Xing and Zhongxu Wei

*Email: 111992026@imu.edu.cn xchen@iphy.ac.cn wji@ruc.edu.cn zhihaicheng@ruc.edu.cn

## Introduction

The well-matched competition between itinerant electrons and polarization has led to the formation of an exotic quantum state - polar metal, the concept of which was proposed by Anderson et al. in 1965 (*1*). From the perspective of meso- and macro-scopic physics, the existence of polar metal is unpromising as the itinerant electrons of metals greatly screen the long-range Coulomb interactions, leading electrical polarization unlikely in bulk materials. However, in some special cases, the weakened conductivity and thus screening capability of itinerant electrons can leave electrical polarizations incompletely compenetrated, forming polar metals (*2*). The first experimental realization of this concept was achieved in $LiOsO_3$ (*3*), subsequently, researchers identified another two-dimensional polar metal $WTe_2$ (*4*).

In addition, polar superconductivity is another special case where two mutually exclusive phases coexist due to weak electron screen. Polar superconductors are rare, but there are still exceptions, in diluted-doped $SrTiO_3$, the heavily screened Coulomb repulsion gives rise to polarized superconducting states, and the superconductivity can be mediated by polarization (*5,6*). The coupled ferroelectricity and superconductivity in bilayer $T_d$-$MoTe_2$ provide a candidate for switching from superconducting to normal metallic states through an external electric field (*7*,*8*). Moreover, polarized superconductivity has been induced via CDW-driven (*9*), pressure (*10*), atomic diffusion (*11*), as well as gating (*12*).

In our previous work, we have investigated a unique quasi-two-dimensional superatomic crystal of $Au_6Te_{12}Se_8$ (ATS), which exhibited a series of sequential-emerged exotic states, as shown in Fig.1A. As a superatomic material, the ATS crystal consists of the step-by-step (1D-2D-3D) stacked high-symmetric molecule-like cubes (0D, $O_h$) with the "soft glue" of covalent-like quasi-bonding in the triclinic lattice. Although the cleaved layers of its *ab*-plane (fig. S1) exist in the ultra-low-symmetric of *P*2 (*P*211), the symmetry-defined order states can still sequentially emerge through the gradual symmetry-reduction of crystal lattice and cube unit. The ATS undergoes triple-cube CDW (TCCDW) phase transition at 120K, antipolar metallic states at 80K (*13*) and two-dimensional superconducting transition at 2.8K (*14*), accompanied by a series of symmetry breaking of electronic structures. High-pressure measurements on ATS display the competition between superconductivity and TCCDW, as well as the reentrant superconductivity induced by pressure (*15*). However, it is still unclear whether the preformed CDW and antipolar charge order still exist or not, and how they interplay and compete with the superconducting quantum state to realize a possible polar superconductor.

**In this work**, we have experimentally investigated the interplay and competition of preformed CDW and antipolar charge order with the emergent superconducting state in the metallic crystal of ATS using an ultra-low-temperature scanning tunneling microscope together with macroscopic transport and Raman measurements. The temperature-dependent preformation (~9.88K) and coherent condensation (~2.8K) of Cooper pairs are experimentally identified. The pre-existent TCCDW is gradually suppressed by the preformed Cooper pairs, and the antipolar charge order is further partially suppressed into a ferrielectric-like polar order by the condensed Cooper pairs of superconducting state. The exotic charge-polarized superconducting state emerged from its delicate interplay and competition with the preformed TCCDW and antipolar charge order, and will inspire more investigation of its potential exotic superconducting properties.

## Results

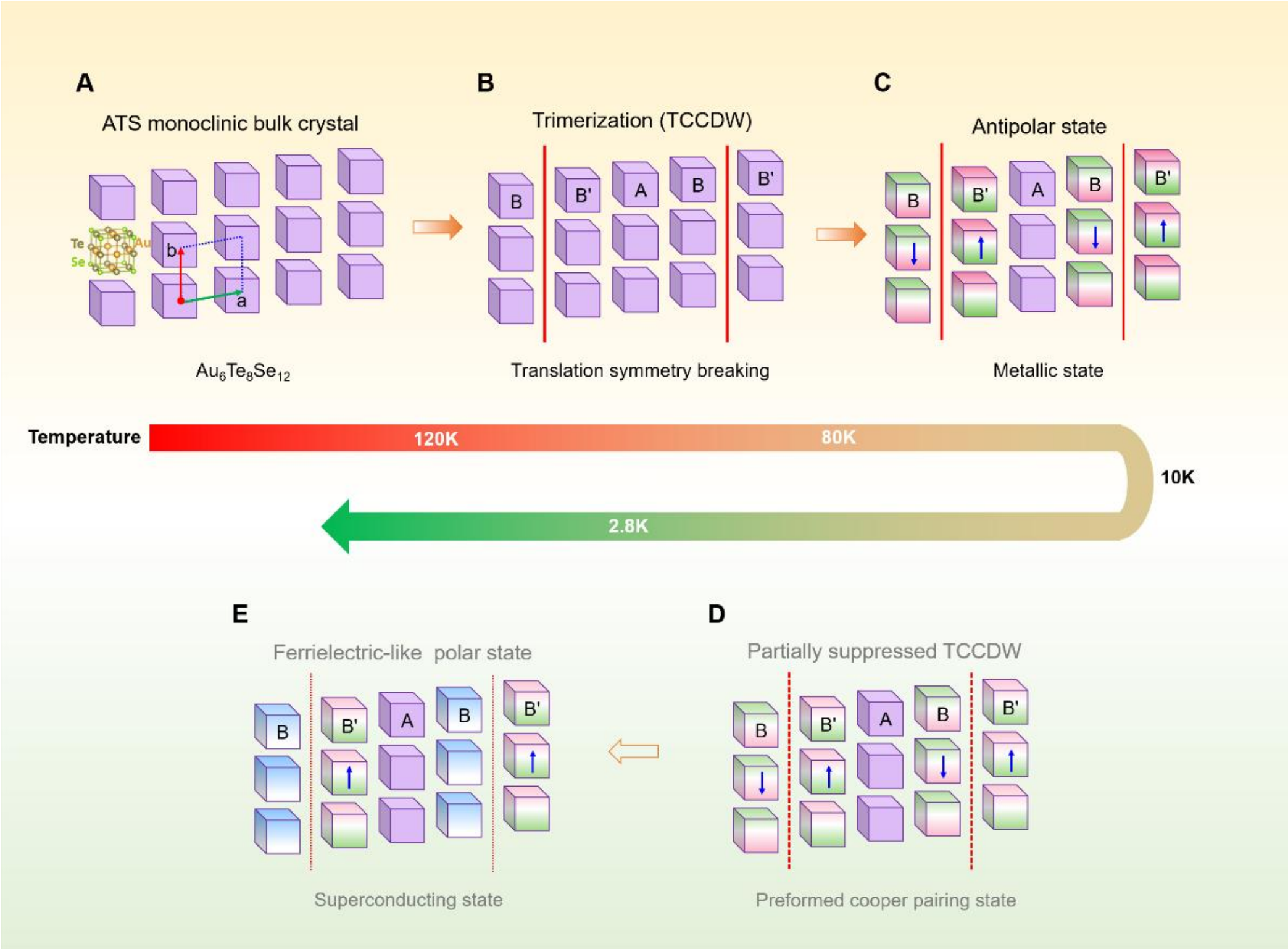

**Figure 1. Emergent ordering states in the superatomic crystal of $Au_6Te_{12}Se_8$ (ATS) as the temperature decreases.** (**A**) The structure of a single cube of ATS is regarded as a superatom (0D). Au, Te, and Se atoms are represented by golden, brown, and green balls, respectively. (**B**) Antipolar metallic states emerge at ~80K with inversion-symmetry-breaking inside the ATS cubes on B and B' chains. (**C**) Trimerization of B, B'and A chains take place at ~120K, which break translation symmetry and give rise to the triple-cube-period CDW (TCCDW). TCCDW continued to strengthen until 10K, and then gradually weakened due to the performed Cooper pairing effect that moved away some electrons (**D**). When the temperature drops below 2.8K, antipolar metallic states are reshaped by coherent superconducting states and transform into a ferroelectric like polar order (**E**). Green and pink filling represent the occupied and unoccupied states of the ATS cube after symmetry breaking, respectively. Purple filling represents the cube that preserves inversion symmetry. The blue filling represents the part of the cube where inversion symmetry is broken but no polarization is formed. The red solid line corresponds to the enhanced TCCDW, while the red dashed line corresponds to its weakening. And the arrow indicates the emergence of new quantum states due to the breaking of electronic states as the temperature decreases.

**Interplay and competition among CDW, antipolar, and superconducting state.**

As a typical macroscopic quantum state, electron pairing into Cooper pairs and subsequent formation of their long-range phase coherence are two essential prerequisites for realizing superconductivity. In conventional superconductors, such as Pb, the paring and coherence occur at the same temperature due to the high superfluid density (*16*). While the electron pairing temperature may become higher than the phase coherence temperature, giving rise to the pre-formation of Cooper pairing above the superconducting critical temperature $T_c$, which have been observed in unconventional cuprate and iron-based superconductors (*17*,*18*). Such unconventional behavior was also discovered in the two-dimensional superconducting state of ATS crystals, as shown in (fig. S2). The pre-existent TCCDW state and antipolar charge order are gradually and separately suppressed and modified by the formation of Cooper pairs and their following long-range phase coherence in the lower temperatures, respectively, as schematically illustrated in Fig. (1D and 1E). The TCCDW state is partially suppressed from ~ 9.88 K by the formation of Cooper pairs, and then the antipolar charge order is selectively modified into a ferrielectric-like polar order due to the further the condensation of Cooper pairs at ~ 2.8 K.

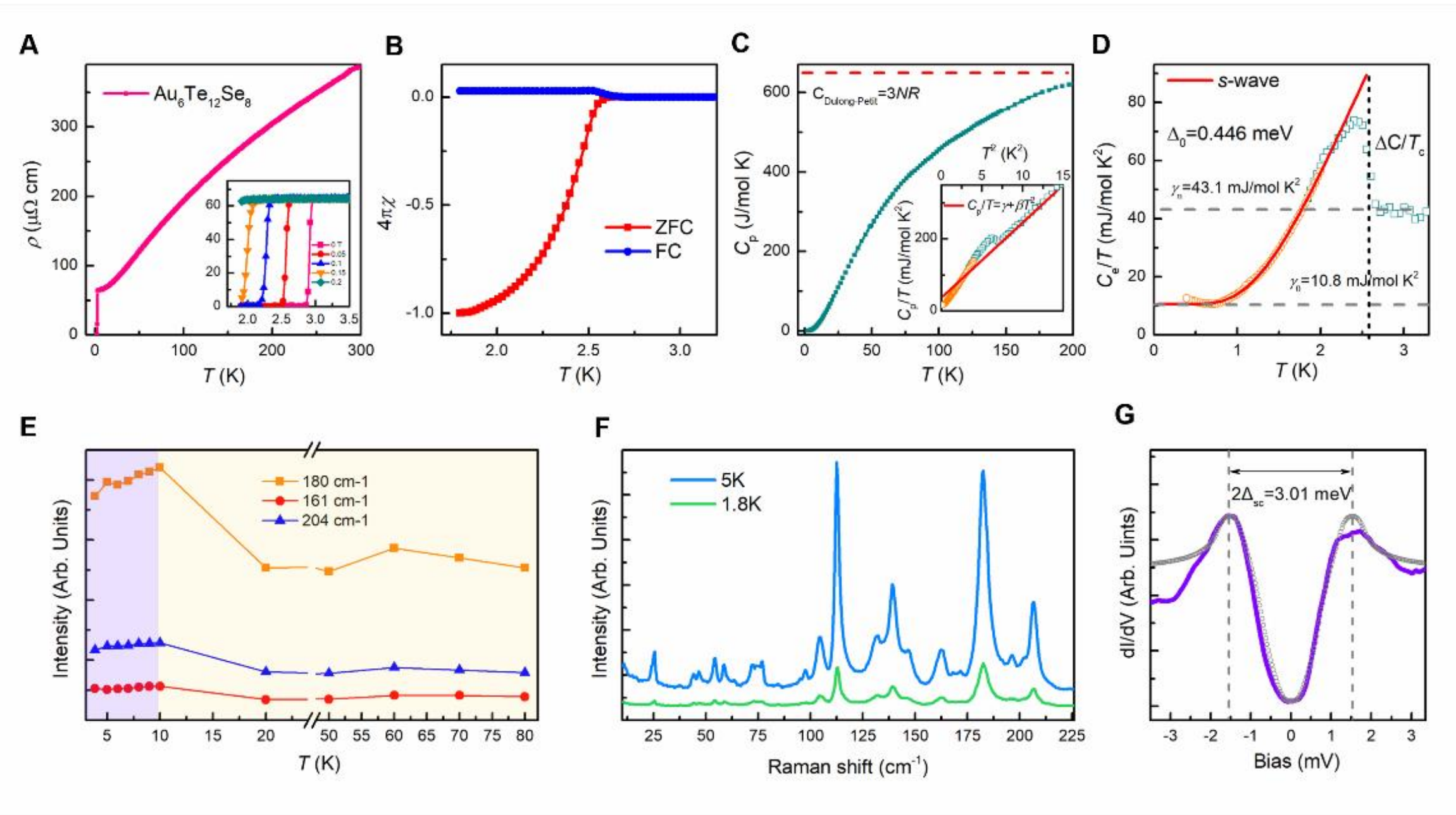

**Figure 2**. **Emergent superconducting properties of $Au_6Te_{12}Se_8$ (ATS).** (**A**) Electrical resistivity of ATS from 1.8 to 300 K. The inset shows the electrical resistivity under different magnetic fields. (**B**) Superconducting volume fractions of ATS under H = 10 Oe. (**C**) Heat capacity ($C_p$) of ATS as a function of temperature, in which the red dashed line represents the Dulong–Petit limit. The inset shows $C_p/T$ versus $T^2$, in which the red straight line represents the fit with the formula $C_p/T = \gamma + \beta T^2$ for the normal-state data. (**D**) Electronic specific heat divided by temperature $C_e/T$ of ATS in the superconducting state, where $C_e = C - \beta T^3$, and the red line is the theoretical curves calculated as the BCS model. (**E**) The temperature-dependent intensity of Raman peaks at 161 cm$^{-1}$, 180 cm$^{-1}$ and 204 cm$^{-1}$. The purple-shaded area marked the temperature-dependent suppressive effect on Raman intensity below 10K. (**F**) Comparison of Raman shifts above (5 K, blue) and below (1.8 K, green) the SC transition temperature ($T_c$ = 2.8 K). The Raman intensity was greatly suppressed at the SC state. (**G**) No clear quasigap of TCCDW was observed around the Fermi level, while a "U-shaped" SC gap of ~ 1.5 mV was clearly distinguished.

**Superconducting properties of ATS crystals.** Figure 2A shows resistance-temperature ($\rho-T$) curves of an ATS crystal under varying magnetic fields. Resistivity $\rho$ decreases upon cooling, exhibiting a metallic characteristic. The superconducting $T_c$ is approximately 2.8 K, as defined by the highest temperature exhibiting zero resistivity ($\rho$ = 0), which is consistent with the literature (*14*). A sharp superconducting transition with a transition width $\Delta T_c$ ~ 0.1K indicate the high quality of the ATS crystal. As shown in the inset of Fig. 2A, low-temperature resistivities of the ATS crystal are measured in magnetic fields up to 0.2 T. Critical temperature $T_c$ decreases with respect to an increasing magnetic field at a rate of ~ 0.5 K/kOe. Figure 2B plots the temperature-dependent electronic magnetic susceptibility ($\chi$–T) of the ATS crystal in both zero-filed cooling (ZFC) and field-cooling (FC) modes under 10Oe. The pronounced diamagnetic signals are exhibited below 2.6 K. From ZFC $\chi$–T curve, the superconducting volume fraction is estimated to be 100% at 1.8 K, confirming its bulk superconductivity.

The specific heat ($C_p$) of the ATS crystal was measured from 0.4 K to 200 K, which is plotted in Fig. 2C. The inset of Fig. 2C shows the plot of $C_p/T$ versus $T^2$. The SC state of the ATS crystal is confirmed by a large superconducting jump at approximately 2.6 K (6.8 $K^2$) in the $C_p/T$ plot, coinciding with the electric resistivity and electronic magnetic susceptibility data. In a normal state, the $C_p$ curve is well fitted by $C_p/T = \gamma + \beta T^2$ from 3–10 K, where the first and the second terms correspond to the normal-state electronic and phonon contribution, respectively. Based on this plot, the Sommerfeld coefficient was fitted to be $\gamma$ = 43.1(1) mJ $mol^{-1}$ $K^2$ and $\beta$ = 20.7(1) mJ $mol^{-1}$ $K^4$. Extrapolating the data to 0 K leads to a residual $\gamma_0$ =10.8(1) mJ $mol^{-1}$ $K^2$, indicating a 1/4 volume contribution from a non-superconducting phase. If $\gamma_s$ is taken as the difference between $\gamma$ and $\gamma_0$ ($\gamma_s = \gamma - \gamma_0$ = 32.3 mJ $mol^{-1}$ $K^2$), the calculated value of $\Delta C_e/\gamma_s T_c$ was derived to be 1.45, see Fig. 2D. This value is close to the Bardeen–Cooper–Schrieffer (BCS) value (1.43) for superconductors in the weak coupling limit. The $C_e/T$ data can be fitted by the expression from the BCS theory, $C_e/T \propto e^{-\Delta/k_B T}$. Furthermore, the fitting yields $2\Delta(0)/k_B T_c$ = 3.95, which is slightly larger than the value of 3.52 for the BCS weak coupling limit. The yielding superconducting gap $\Delta(0)$ = 0.446 meV, which represents the condensed energy of the Cooper pairs in the ATS crystal that undergoes complete long-range coherence. The good agreement between the measured data and the BCS fitting evidences a gap of an *s*-wave isotropic superconductor.

**Raman and STS measurements.** Temperature-dependent Raman shift spectra were measured at a series of temperature from 80 K to 3.85 K, which exhibit six primary feature peaks (see Supplementary fig. S3). We plotted the peak intensities for three of them, residing at 180 $cm^{-1}$, 161$cm^{-1}$, and 204 $cm^{-1}$, in Fig. 2E a turning point of 9.88 K is identified in the intensity plots, as the peak intensities gradually decrease from 10 to 3.85 K before the superconducting transition. In comparison to those measured at 5 K, the SC state significantly suppresses the intensities of Raman peaks acquired at 1.8 K, as shown in Fig. 2F Such characteristic temperature of 9.88 K is identified again in STS measurements. Figure 2G shows STS spectra acquired at ~ 300 mK, with a clear "U-shaped" SC gap. Its size (Δ) is defined as half of the energy spacing between two coherence peaks and is estimated to be approximately 1.50 meV, which is over three times the superconducting gap size ($\Delta_0$= 0.446 meV) as determined from the electronic specific heat measurement Fig. 2D. According to Ref. (*19*,*20*), the STS-measured large SC gap ($\Delta_{SC}$ = 1.50 meV) corresponds to a nominal transition temperature of approximately 9.88 K (Numerical fitting using BCS model (*21*)), which is much larger than the transport-measurement-determined SC temperature of approximately 2.8 K. Therefore, we assess this characteristic temperature of 9.88 K as the pre-forming temperature of Cooper pairing.

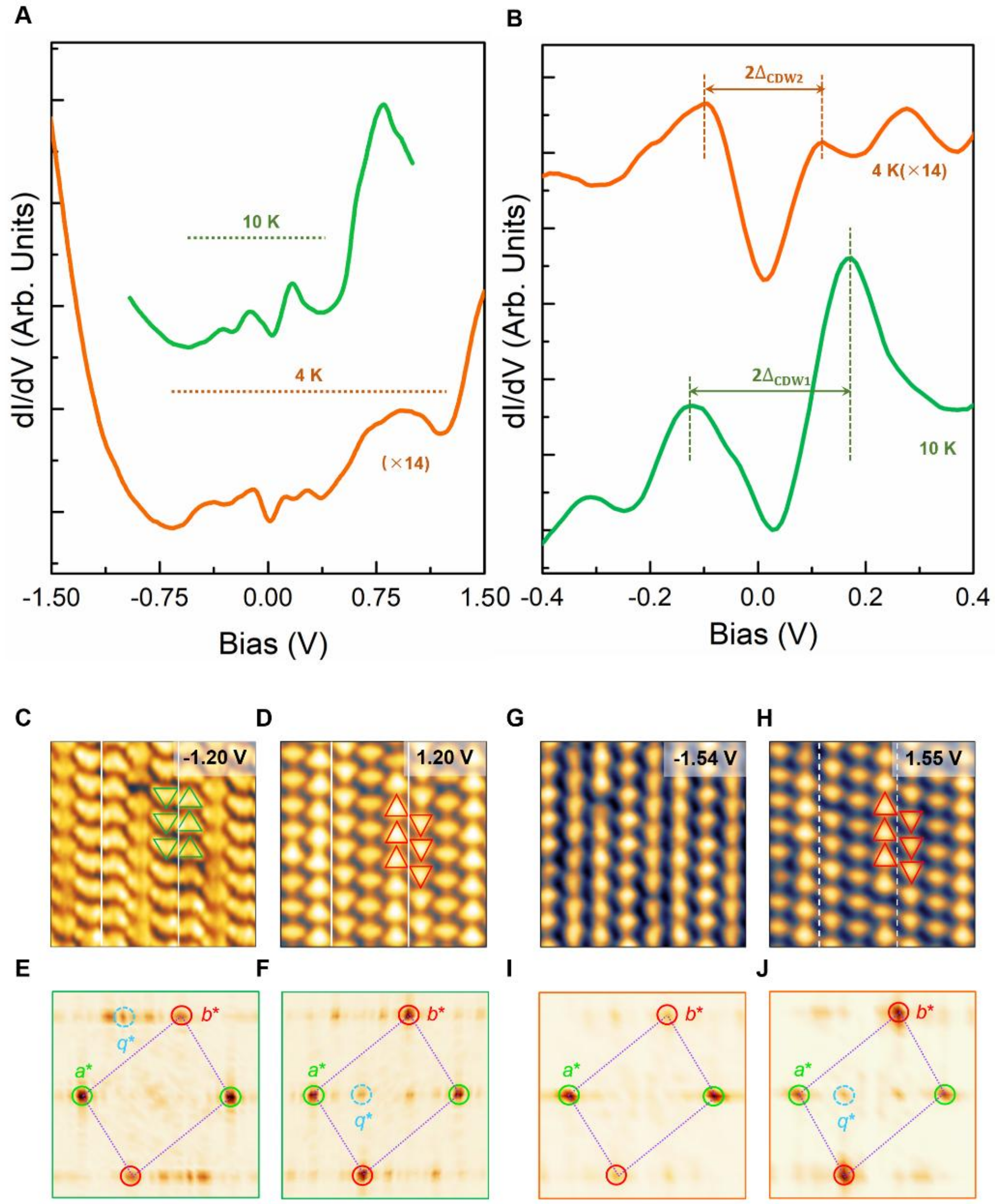

**Figure 3**. **TCCDW states of ATS crystal at ~ 10 K and ~ 4 K.** (**A**) STS spectra of ATS crystal at 10 K (green) and 4 K (orange), respectively. The molecular-like gaps increased from ~0.9 eV (green dotted line) to ~ 1.9 eV (orange dotted line) with decreased in-gap states. (**B**) The observed quasigap around the Fermi level is decreased from 0.17 eV (10 K) to 0.11 eV (4 K, inset spectra), coincide the suppressed TCCDW state in Fig. 2A. (**C,E**) STM topography (**C,D**) and corresponding FFT pattern (**E,F**) of ATS at ~ 10 K. (**G,J**) STM topography (**G,H**) and corresponding FFT pattern (**I,J**) of ATS at ~ 4 K. The FFT spots (blue circles) of TCCDW can only be clearly resolved at the unoccupied states with the underlying Bragg spots (the green circles correspond to $a^*$ and red circles correspond to $b^*$) and the marked white lines.

**Suppression of the TCCDW order after electron pairing**. Figure 3A shows d*I*/d*V* spectra comparatively obtained at 10 K and 4 K, which represent density of states (DOSs) on the surface of the ATS crystal. The intensity around the Fermi level of the 4K d*I*/d*V* spectrum is significantly decreased from that of 10 K. Additionally, the size of the quasi-gap of the TCCDW order is decreased from 0.17 eV (acquired at 10 K), to 0.11 eV (measured at 4 K), as shown in Fig. 3B. All of these spectra features indicate that the pre-existent TCCDW state is gradually suppressed by formation of Cooper pairs initialized from 9.88 K, consistent with the temperature-dependent Raman measurements shown in Fig. 2E.

Figure (3C and 3D) show an occupied- and an unoccupied-state STM topographic images acquired on the surface of the ATS crystal at 10 K, in which the tripe-cube-period characteristics of TCCDW are clearly resolved consistent with the corresponding FFT patterns in Fig. (3E and 3F). The interweaved antipolar charge order and TCCDW state (highlighted by the dashed lines) are highlighted by the distinct "up-triangular" (UT) and "down-triangular" (DT) shapes at the occupied state, as well as "down-triangular" (UT) and "up-triangular" (DT) shapes at the unoccupied state. Fig. (3G and 3H) shows two typical STM topographic images for both occupied and unoccupied states of the ATS surface acquired at 4 K, in which the tripe-cube-period characteristics can only be resolved at the unoccupied states (denoted with "$q^*$" in blue in Fig. 3J. In comparison with explicit features observed at 10 K, the tripe-cube-period characteristics of TCCDW become substantially blurry in STM topographic images acquired at ~4 K, especially for the occupied states. This trend of blurry tripe-cube-period is confirmed by the corresponding FFT patterns illustrated in Supplementary (fig.S4 and S5). These observations indicate that the pre-existent TCCDW state is gradually suppressed, especially at the occupied states, by the formation of Cooper pairs initializing from 9.88 K.

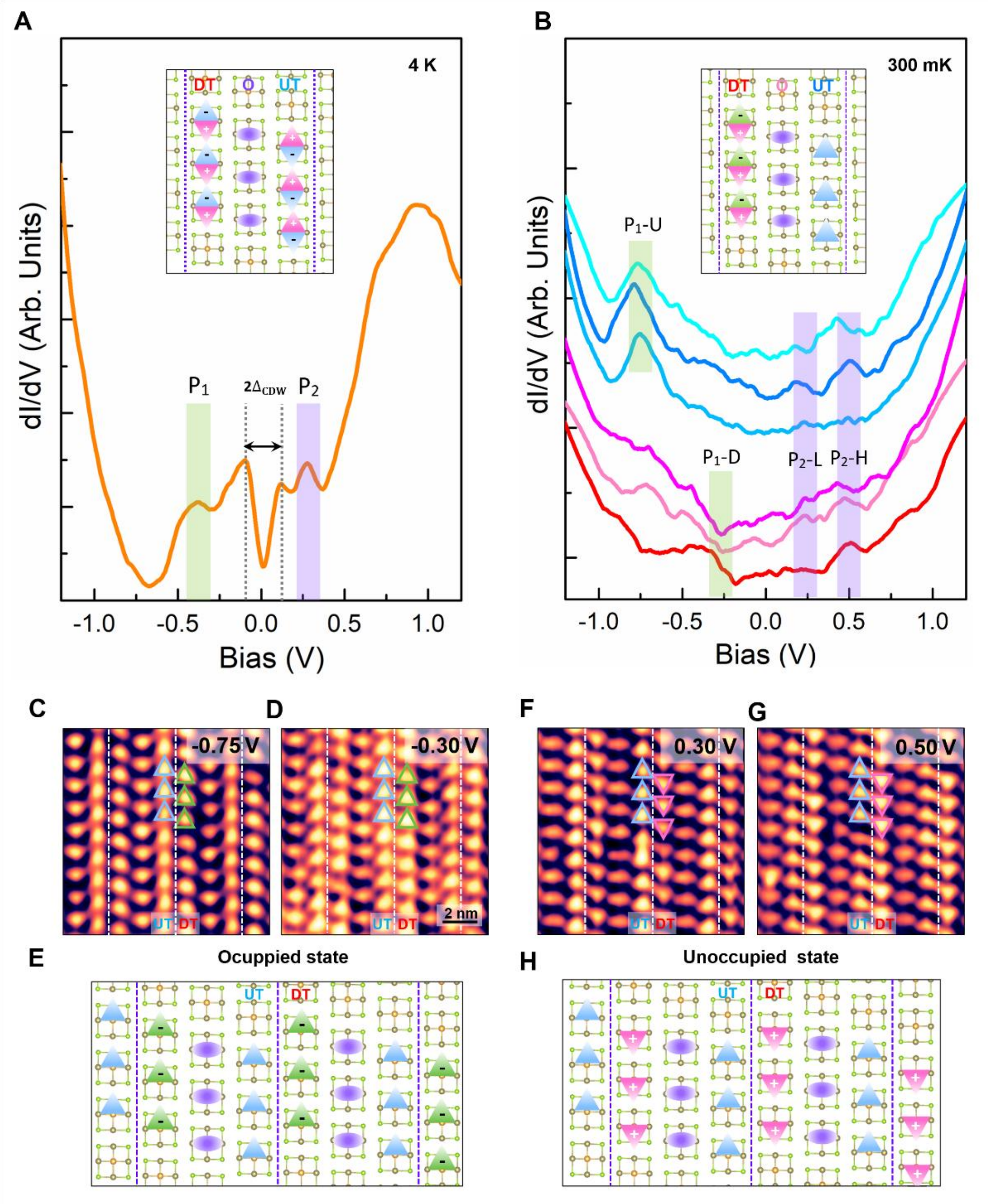

**Figure 4. Emergent ferrielectric-like polar order at the superconducting state of ATS. (A)** STS spectra of the in-gap states at ~ 4 K, which display the characteristics in-gap states of P1 (-0.39 V) and P2 (0.28 V), as well as the CDW quasigap. Inset: Illustration of antipolar charge order within a triple-cube CDW period. **(B)** STS spectra of the in-gap states at ~ 300 mK, which display the chain-specific in-gap states of $P_1$-U (-0.75V) and $P_1$-T (-0.3V), as wells as the $P_2$-L (0.25V) and $P_2$-H (0.5V) states. **(C, D, F, G)** STM topography images obtained around the in-gap states marked in (B). The TCCDW could be readily resolved and marked by the white dashed lines. In the unoccupied states, the inversion-symmetry of UT and DT chains is still preserved and marked by the up- and down-triangles, respectively. While in the occupied states, the inversion-symmetry of UT and DT chains is breaking and marked by the up- and up-triangles, respectively. The preserved and breaking inversion-symmetry of UT and DT chains is in agreement with the STS spectra of Fig. 4B. **(E, H)** Illustration of in-gap polarized electronic states at filled (E) and empty (H) states. The preformed antipolar order has been converted to a ferrielectric-like polar order by the emergent SC state.

**Ferrielectric-like polar states coexisting with the SC state.** Figure 4A shows a representative high-resolution d*I*/d*V* spectrum acquired at ~4 K, which displays characteristic peak-like states of $P_1$ (-0.39 V, in green) and $P_2$ (0.28 V, in violet), as well as a CDW quasi-gap. Although the spectra weight (DOS) around the Fermi level is greatly suppressed by the pre-formed Cooper pairs before the SC transition, the interweaved TCCDW and antipolar charge orders can still be resolved in STM tomographs at this temperature. In particular, the inversion symmetry of DT and UT chains is nearly preserved with respect to O chains. Figure 4B shows a series of representative chain-specific d*I*/d*V* spectra measured at ~300 mK. This temperature is far below $T_c$ and the sample is already in its SC state. From these spectra, no clear quasi-gap can be identified around the Fermi level, as the TCCDW order is greatly suppressed. Moreover, the energy degeneracy of UT and DT cubes is further lifted that the $P_1$ state (-0.39 V) splits into $P_1$-U (-0.75 V, green shadow on cold-colored curves) and $P_1$-D (-0.30 V, on warm-colored curves) states, residing around the UT and DT chains, respectively. Differently from the occupied states, the unoccupied $P_2$ (~0.28 V) state is only energetically shifted and split into the $P_2$-L (~0.25 V) and $P_2$-H (~0.50 V). These two states can be detected at both the UT and DT chains, exhibiting no spatial splitting. It is noted that the energetical and spatial splitting of UT and DT chains mostly occurs at the occupied states (see supplementary information fig. S6 and S7).

STM imaging of these four split $P_1$ and $P_2$ states further illustrate the broken spatial inversion-symmetry of UT and DT chains. Figure (4C and 4D) show STM topography images for the $P_1$-U and $P_1$-D states, in which the triple-cube periods of TCCDW can still be readily resolved along the *a*-axis, however, the previously observed UT and DT pairs no longer exist. For states $P_1$-U and $P_1$-D, both UT and DT chains exhibit "up-triangular" shapes, as highlighted by the overlaid blue and green triangles Fig. (4C and 4D) and schematically illustrated by a model shown in Fig. 4E. For the unoccupied $P_2$-L and $P_2$-H states, Fig. (4F and 4G) depict their STM topographic images. At both states, the UT (DT) chains always exhibit "up- (down-) triangular" shapes, as highlighted by the overlaid blue and pink triangles Fig. (4E and 4H) and schematically illustrated in a model shown in Fig. 4H (see Supplementary fig. S8 for more details).

## Discussion

The polar characteristic of the UT and DT chains within the pre-existing antipolar charge order persists across, rather than being suppressed by, the superconductivity transition. It is, however, reshaped by the emergence of phase coherence in the SC states. On the *ab*-cleaved surface of an ATS crystal, identical superatomic ATS cubes are stacked face-to-face along the ***b*** axis, with multiple chains running parallel to form the two-dimensional surface lattice. Within each chain, adjacent cubes interact primarily through four covalent-like Te···Te quasi-bonds, giving rise to strong overlap of Te *p* orbitals and highly degenerate electronic states and thus high density of states at the Fermi level. In the antipolar phase, interactions among these states lift the local $C_{4z}$ rotational symmetry of each cube, yielding a triangular charge density motif projected onto the top surface. Within each tc -CDW period, two adjacent chains display triangles of opposite orientations, establishing an antisymmetric (antipolar) charge-ordering pattern that can minimize real-space Coulomb repulsion (Fig. 4A).

Upon entering the superconducting phase, electron pairing enhances the effective attraction among electrons and opens a gap at the Fermi level. The local $C_{4z}$ symmetry of each ATS cube remains broken and the triangular motif persists, however, the mechanism for minimizing Coulomb repulsion changes fundamentally. Rather than appearing in alternating orientations, the degeneracy of the electronic states is lifted in both energy and real space that a state at a given energy localizes on an individual chain with the tc-CDW period, while triangles associated with different energies align in the same direction. This alignment also effective suppresses local charge variations, favoring the formation of the coherent superconducting state. Because the empty bands remain unoccupied, they do not compete with electron pairing. The onset of superconductivity thus leaves the (anti)polar configuration of the unoccupied states essentially unaffected, with only a modest energy splitting observed from 0.28 eV to 0.25 eV and 0.50 eV. This splitting is, most likely, a consequence of the chain selective lifting of degeneracy in the corresponding occupied bands. These findings reveal that the charge polarization in ATS crystals derives not only from Fermi-surface electrons but also from valence electrons residing at lower energies, mediated by the Te…Te and other quais-covalent interactions. As a result, the superconductivity coexists with, rather than suppresses, the charge polar state, and subtly reshapes its microscopic texture from anitpolar to unipolar.

In summary, we identified a concise temperature sequence for ATS (fig. S9). As we revealed previously, the tc-CDW state emerges at approximately 120 K through a transverse trimerization along the $a$ axis. Around 80 K, this lattice modulation locks antipolar electronic states onto the two outer (B, B′) chains, forming an antipolar metal state. In this work, upon further cooling, Cooper pairing is evident at 9.88 K and superconductivity appears at 2.8 K. The formation of Cooper pairs first weakens the tc-CDW below 9.88 K and then their coherent condensation partially converts the antipolar charge order into a unipolar pattern. These results uncover a ground state where tc-CDW, unipolar charge order and superconductivity coexist, intricately interwind, and subtly compete (see Supplementary fig. S10 for more details). Our STM/STS imaging resolves the polar charge texture, while both local spectroscopy and bulk measurements confirm the emergence of superconductivity. To the best of our knowledge, ATS is the first material in which superconductivity can coexist and stabilize a unipolar metal in an antipolar metal parent phase. The ultra-low-symmetry and strong interactions in the cubic superatomic ATS lattice offers a unique platform to probe polarization-, chain- and cube-specific superconducting parameters, an arrangement reminiscent of, but distinct from, pair-density-wave physics. Moreover, more careful and higher sensitivity transport, magnetic, and optical measurements, such as magnetic susceptibility, SHG, ESR, should clarify the anisotropic properties of the coexistent ordering states. More broadly, the chemical tunability of superatomic crystals, such as another endohedral clusters as building blocks, soft inter-block interactions, and engineerable cluster and lattice symmetries, opens an avenue to designing further quantum states in superatomic materials.

## Methods

**Sample preparation.** Single crystals of $Au_6Te_{12}Se_8$ (ATS) were grown using the self-flux method, as described in ref (*14*). Starting materials with high purity Au powder (99.99%, Sigma Aldrich), Te powder (99.99%, Sigma Aldrich), and Se powder (99.99%, Sigma Aldrich) are stoichiometrically weighted and sealed in an evacuated silica tube in high vacuum and subsequently mounted into a muffle furnace. The furnace was heated up to 850 °C in 24 h and dwelled 48 h. Afterward, the furnace was slowly cooled down to 450 °C in 7 days and then shut down.

**STM measurements.** STM experiments were performed in a commercial ultrahigh vacuum STM system (USM-1300-$^3$He system with a 16-T magnet) operated in the Synergic Extreme Condition User Facility, Beijing, China. The energy resolution can reach below 0.26 mV. PtIr alloy tips calibrated on clean Ag(111) surfaces were used for all our STM measurements. The STM topography was obtained in the constant-current mode, and the differential conductance (*dI/dV*) spectra and maps were acquired using a standard lock-in technique at a frequency of 879.9 Hz with modulation voltages of 20 mV corresponding to Fig. 3(C-F), Fig. 4(C and D), Fig. 4(F and G), 0.6 mV to Fig. 2G, respectively. The STM measurements were respectively performed at 4.6K and 300mK to obtain the real space electronic structure above and below the ATS superconducting transition temperature ($T_c$~2.8 K). The STM/STS data were processed using Gwyddion and WSxM software.

**XRD, SEM, transport, and Raman measurements.** The microstructure of ATS was examined using a scanning electron microscope (SEM, SU5000, HITACHI). The chemical composition of ATS was determined by the Energy Dispersive Spectrum (EDS). Powder X-ray diffraction (PXRD) patterns in the *ab*-plane of single crystal ATS are measured using a Rigaku SmartLab 9 kw diffractometer with the Cu-Kα anode (λ = 1.5408 Å). The electrical resistivity (ρ), V-I curves, and specific heat capacity (Cp) of ATS were measured through the standard four-wire method and the thermal relaxation method using the physical property measurement system (PPMS-16, Quantum Design). The specific heat below 1.8 K was measured in the physical property measurement system with a $He^3$ inset. The dc magnetic susceptibility (χ) was characterized using SQUID (MPMS, Quantum Design). The temperature-dependent Raman was measured in the Micro confocal Raman spectrometer (LabRAM HR Evolution, HORIBA) with a continuous-helium-flow optical cryostat.

## Acknowledgments

This project is financially supported by the National Natural Science Foundation of China (Grant Nos. 52302010, 52250308, 92477128), the Ministry of Science and Technology of China, National Key Research and Development Program "Physical Regulation" Special Project (No. 2023YFA1406500), Major research project of China (No. 92477128), Natural Science Foundation of Inner Mongolia Department of Science and Technology Autonomous Region Youth Fund No. 2024QN01010), High level talent introduction and research funding support of Department of Human Resources and Social Security of Inner Mongolia Autonomous Region (No. 12000-150422225). This work was supported by the Synergetic Extreme Condition User Facility (SECUF, https://cstr.cn/31123.02.SECUF).

**Author contributions**

Z.C., S.X., X.C., and W.J. conceived the research project. S.X., Z.W., J.Z., Z.Y., J.Z., and T.Q. performed the STM experiments and analysis of STM data. X.C., T.Y., J.G., and H.C. grew the single crystals. X.C., T.S., H.R., M.C., and H.L. performed transport, SEM, and XRD measurements. X.C., S.X., F.J., H.C., and S.Z. performed Raman measurements. X.L., W.Z., and W.J. conducted a theoretical analysis of the experimental data. S.X., X.C., and Z.C. wrote the manuscript with inputs from all authors.

**Competing Interests**

The authors declare no competing financial interests.

**Data Availability**

The authors declare that the data supporting the findings of this study are available within the article and its Supplementary Information.